\documentclass[journal=apchd5,manuscript=article]{achemso}

\usepackage[version=3]{mhchem}
\usepackage[squaren,Gray]{SIunits}
\usepackage{graphicx}
\usepackage{amsmath}

\newcommand*\textGamma{$\mathrm{\Gamma}$}

\author{K. Guilloy}
\author{N. Pauc}
\email{nicolas.pauc@cea.fr}
\author{A. Gassenq}
\author{Y.M. Niquet}
\author{J.M. Escalante}
\author{I. Duchemin}
\author{S. Tardif}
\affiliation{Universit\'e Grenoble Alpes, F-38000 Grenoble, France \\CEA, INAC, F-38000 Grenoble, France}

\author{G. Osvaldo Dias}
\author{D. Rouchon}
\author{J. Widiez}
\author{J.M. Hartmann}
\affiliation{CEA-LETI, Minatec Campus, F-38054 Grenoble, France}

\author{R. Geiger}
\author{T. Zabel}
\author{H. Sigg}
\affiliation{Laboratory for Micro- and Nanotechnology (LMN), Paul Scherrer Institut, CH-5232 Villigen, Switzerland}

\author{J. Faist}
\affiliation{Institute for Quantum Electronic, ETH Zurich, CH-8093 Zurich, Switzerland}

\author{A. Chelnokov}
\author{V. Reboud}
\affiliation{CEA-LETI, Minatec Campus, F-38054 Grenoble, France}

\author{V. Calvo}
\affiliation{Universit\'e Grenoble Alpes, F-38000 Grenoble, France \\CEA, INAC, F-38000 Grenoble, France}

\title{Germanium under high tensile stress: nonlinear dependence of direct band gap vs. strain}


\begin{document}

\newpage

\begin{abstract}

Germanium is a strong candidate as a laser source for silicon photonics. It is widely accepted that the band structure of germanium can be altered by tensile strain so as to reduce the energy difference between its direct and indirect band gaps. However, the conventional gap deformation potential model most widely adopted to describe this transition happens to have been investigated only up to \unit{1}{\%} uniaxially loaded strains. In this work, we use a micro-bridge geometry to uniaxially stress germanium along [100] up to $\varepsilon_{100} = \unit{3.3}{\%}$ longitudinal strain and then perform electro-absorption spectroscopy. We accurately measure the energy gap between the conduction band at the \textGamma{} point and the light- and heavy-hole valence bands. While the experimental results agree with the conventional linear deformation potential theory up to \unit{2}{\%} strain, a significantly nonlinear behavior is observed at higher strains. We measure the hydrostatic and tetragonal shear deformation potential of germanium to be $a = \unit{-9.1 \pm 0.3}{\electronvolt}$ and $b = \unit{-2.32 \pm 0.06}{\electronvolt}$ and introduce a second order deformation potential. The experimental results are found to be well described by tight-binding simulations. These new high strain coefficients will be suitable for the design of future CMOS-compatible lasers and opto-electronic devices based on highly strained germanium.

\end{abstract}

\newpage

\section{Introduction}

Despite its indirect band gap, germanium is a promising candidate as a light emitter for the integrated silicon photonics. The application of several percent of tensile strain has been proposed to increase its optical emission efficiency. Since it reduces the energy difference between its direct and indirect band gaps, it could in principle lead to laser emission \cite{Liu07}.

Significant efforts have been undertaken in the band gap engineering of germanium in order to overcome this energy difference. Applying high amounts of tensile strain in Ge should lead to a direct band gap behavior \cite{Boucaud13, Boztug14}. Various techniques have been proposed to reach such high strain levels, including the use of silicon nitride as an external stressor \cite{Kersauson11, Jain12, Ghrib13, Capellini14}, the stretching of germanium nanowires \cite{Greil12, Guilloy15}, as well as the mechanical amplication in a micro-bridge of the small tensile strain present in a germanium layer grown on a silicon substrate \cite{Suess13, Nam13, Gassenq15}.

The behavior of the direct band gap under large hydrostatic compressive stress is known to be non linear \cite{Welber77, Ahmad89, Goni89, Li94}. However, for the range of tensile strain which is of interest for the exploration of direct band gap germanium, in the order of several percent, the current theory is based on first order coefficients known as deformation potentials that are experimentally confirmed only up to \unit{1}{\%} strain for [100] uniaxial stress \cite{Balslev67, Chandrasekhar77, Pollak68, Hall62}.  In particular, the most commonly cited deformation potentials predict an indirect-to-direct band gap transition at \unit{4.4}{\%} \cite{VandeWalle89} leading to claims about the direct nature of the band gap based only on strain measurements \cite{Sukhdeo14}.

In this work, we study experimentally the effects of strong uniaxial tensile stress along [100] in Ge micro-bridges using optical electro-absorption measurements. Based on the Franz-Keldysh effect, the electro-modulation techniques have proven to be more accurate than photoluminescence for the measurement of optical transitions in semiconductors such as germanium \cite{Ghosh68, Hamakawa68}, particularly in the transmission configuration \cite{Frova66}. Using this technique, we demonstrate a significant nonlinearity in the strain-band gap relation.

We then compute the effects of strain on the band structure of germanium using a tight-binding model and compare this theoretical work to the spectroscopic measurements. We finally extract second-order deformation potentials from both theoretical and experimental data.

\section{Electro-absorption spectroscopy}

\begin{figure}[ht]
\centering
\includegraphics{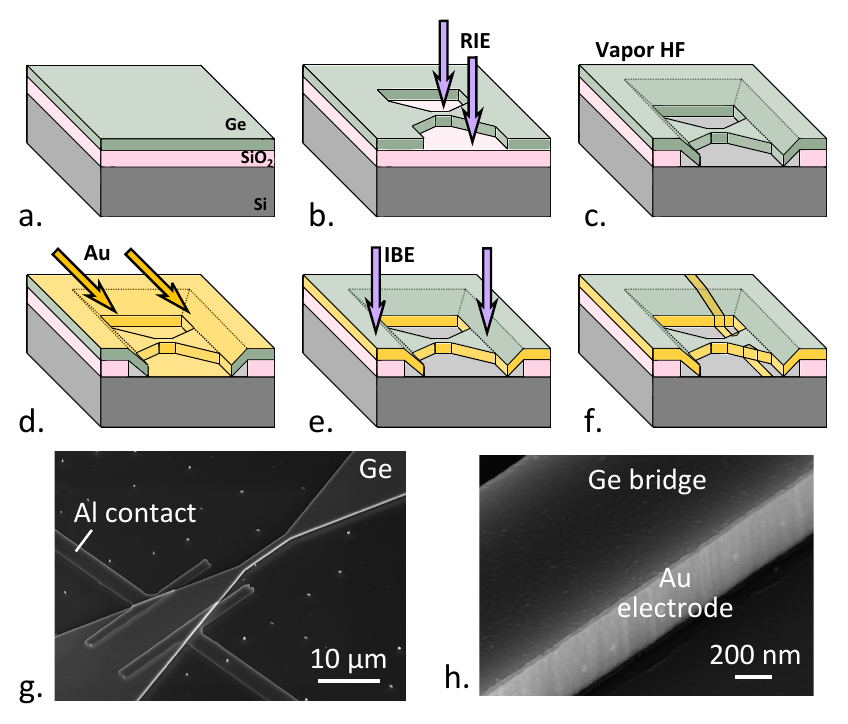}
\caption{\label{fgr:process}Fabrication process of the strained germanium samples for electro-absorption. (a) Germanium-On-Insulator (GeOI) initial substrate. (b) Reactive Ion Etching of the micro-bridge. (c) Vapor HF etching of the silicon dioxide sacrificial layer. (d) All around gold deposition at \unit{45}{\degree}. (e) Ion Beam Etching of the top gold layer. (f) Contacting of the lateral electrodes with aluminum deposition. (g) Scanning Electron Microscopy (SEM) image of such a device, showing the micro-bridge, its stretching arms and the aluminum contacts. (h) Close-up SEM image of the lateral gold layer on a micro-bridge.}
\end{figure}

Using the method described by Suess et al.\cite{Suess13}, we fabricate highly strained germanium micro-bridges from germanium-on-insulator layers (figure \ref{fgr:process}.a to c). These structures, normally suspended in air, are brought into contact with the silicon substrate, allowing further technological steps (see method section).

In order to perform electro-absorption spectroscopy experiments, an electric field must be applied to the strained germanium sample. For this purpose, \unit{60}{\nano\meter} of gold was deposited at \unit{45}{\degree} with a planetary rotation of the sample in order to cover all the sides of the microbridges (figure \ref{fgr:process}.d). An argon Ion Beam Etching (IBE) step removed the top layer of gold, leaving about \unit{20}{\nano\meter} of gold only on the side walls of the patterned bridge structures (figure \ref{fgr:process}.e). Finally, deep UV lithography was employed to pattern the contacts, followed by the deposition of \unit{300}{\nano\meter} of aluminum. After lift-off, the gold electrodes on both sides of the micro-bridge are contacted by aluminum contacts (figure \ref{fgr:process}.f), as seen in figures \ref{fgr:process}.g and \ref{fgr:process}.h. Due to its low ultimate tensile stress (\unit{0.1}{\giga\pascal} \cite{Howatson12}) and negligible thickness, the influence of gold on the strain in germanium is estimated to be less than \unit{150}{ppm} and can be disregarded.

\begin{figure}[ht]
\centering
\includegraphics{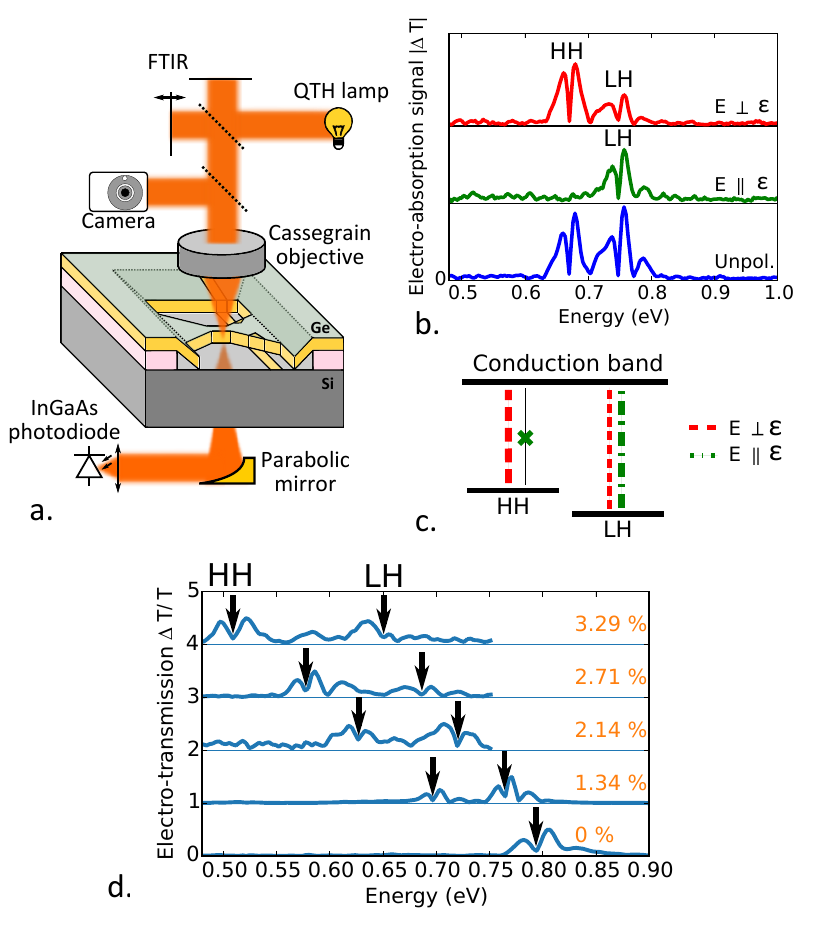}
\caption{\label{fgr:electroabs}
(a) Simplifed schematics of the electro-absorption optical experiment. (b) Experimental electro-absorption signal on a strained micro-bridge under unpolarised light, with a polarisation parallel to the strain axis and with a polarisation perpendicular to the strain axis. (d) Schematics showing the selection rules relative to the transition to the heavy hole band (HH) and the light hole band (LH). The thickness represents the relative momentum matrix elements of the transitions. The HH band exhibits an extinction for light polarised along the stress axis. (e) Electro-absorption signals measured under unpolarised light for increasing strain in micro-bridges.}
\end{figure}

We performed electro-absorption spectroscopy for optical polarisations with an electric field parallel and perpendicular to the stress axis of the micro-bridge. Experimental electro-absorption spectra from strained micro-bridges exhibit two main oscillating components, as seen in figure \ref{fgr:electroabs}.b. We observe an extinction of the lowest energy structure for the polarisation parallel to the bridge axis. This is explained by the selection rules for the valence bands. As shown in figure \ref{fgr:electroabs}.c, the transition from the conduction band at the \textGamma{} point to the heavy hole valence band has a null momentum matrix element for a light polarisation with an electric field along the stress axis, while the vertical transition with the light hole band does not exhibit such an extinction \cite{Chuang12}. It is possible to extract from this spectrum the direct band gaps to the light and heavy hole bands $E_{g,LH}$ and $E_{g,HH}$. For this strained germanium microbridge (at \unit{1.6}{\%}), we measure for instance $E_{g,LH} = \unit{0.75 \pm 0.01}{\electronvolt}$ and $E_{g,HH} = \unit{0.67 \pm 0.01}{\electronvolt}$. Because of the low absorption coefficient of germanium at its indirect band edge, it is not possible to extract the energy of the transitions between the L valley of the conduction band and the valence bands.

By performing electro-absorption spectroscopy on micro-bridges with different stretching arm lengths, it is possible to measure the spectrum as a function of longitudinal strain, as shown in figure \ref{fgr:electroabs}.d. For strain levels above \unit{1}{\%}, we clearly observe the splitting between the light and heavy hole bands. We reached strains up to \unit{3.3}{\%} and the corresponding band gaps are $E_{g,HH} = \unit{0.51}{\electronvolt}$ and $E_{g,LH} = \unit{0.65}{\electronvolt}$.

From a set of 19 strained micro-bridge samples, we finally plotted the position of the band gaps as a function of longitudinal strain in figure \ref{fgr:concl}. We first compared these experimental data points with theoretical curves obtained using deformation potentials from the model-solid theory \cite{VandeWalle89} which are commonly cited for strained germanium (for instance in ref. \cite{Boztug14, Jain12, Greil12}), as plotted in figure \ref{fgr:concl}. They are in very good agreement below \unit{2.0}{\%}, i.e. the deformation potentials describe well the evolution of the direct band gaps with the uniaxial [100] stress. However, experimental transition energies above \unit{2.0}{\%} are systematically lower than the theoretical curves. We therefore conclude that a first order approach using deformation potentials cannot describe precisely the behavior of the band gaps at very high strain and that higher order effects have then to be taken into account.

\section{Comparison with theoretical models}

We have computed the direct band gap of germanium as a function of the uniaxial [100] loading using the tight-binding model of ref. \cite{Niquet09}, adapted to the case of uniaxial stress \cite{Escalante15}. This tight-binding model was built on top of \emph{ab initio} data and designed to reproduce the band structure of germanium under arbitrary strains in the $\pm 5\%$ range. Since this model targets the zero temperature band structure, the calculated heavy- and light-hole bandgaps were corrected for room temperature effects using Varshni's model \cite{Varshni67}.

\begin{figure}[ht]
\centering
\includegraphics[width=8cm]{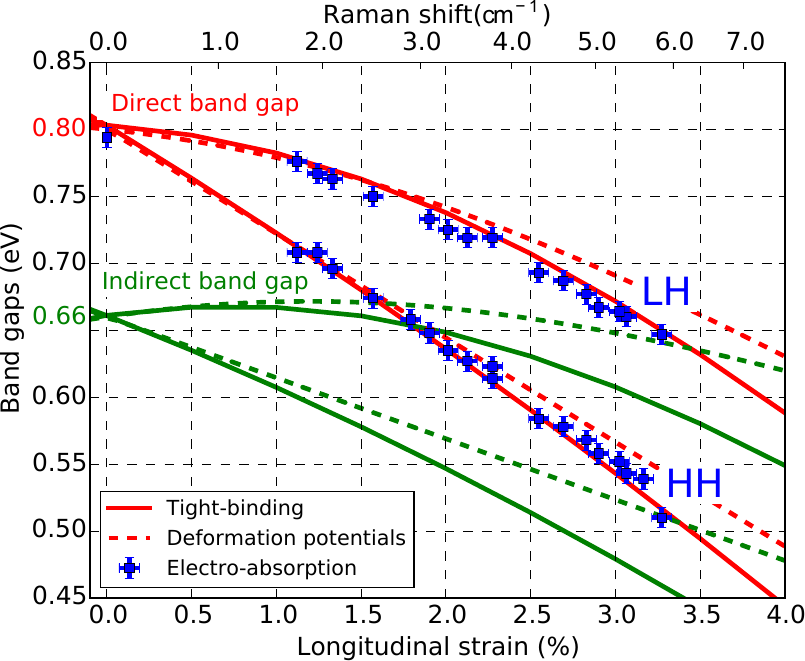}
\caption{\label{fgr:concl} Band gaps measured by electro-absorption (blue squares) as a function of longitudinal strain measured by Raman spectroscopy. These experimental points are compared to theoretical band gap values computed with deformation potentials from \cite{VandeWalle89} (dashed red line) and with a tight-binding model (solid lines). The upper curve corresponds to the transition with the light hole band while the lower one is with the heavy hole band.}
\end{figure}

We plot in figure \ref{fgr:concl} the results of the simulations, as well as the experimental data points and the band gaps calculated with the deformation potentials. Experimental points are in very good agreement with the tight-binding model, which thus properly predict the evolution of the direct band gaps with strain, at least up to \unit{3.3}{\%}.

Additionally, the tight-binding model allows us to compute the evolution of the indirect band gap which cannot be observed in our experiment. We also obtain a significant nonlinearity for this transition, as seen in figure \ref{fgr:concl}. This has a consequence on the position of the crossover point between the indirect and the direct band gap regimes: it is predicted to occur at \unit{5.6}{\%}, instead of \unit{4.4}{\%} for the deformation potential theory \cite{VandeWalle89} as has been previously inferred in ref. \cite{Sukhdeo14}.

To take into account the non-linearity in the relation between the band gap and strain, we introduce an extra term in the conduction band energy shifts. We choose to add a second order $a$ term because the non linearity of the tight-binding conduction band edge is an even function. Furthermore, since the usual deformation potential theory can properly describe the splitting between the light- and heavy-hole bands (both theoretically and experimentally), a second order $b$ term is not required. We also introduce a similar second order term in the expression of the indirect band gap. These second order deformation potentials, only valid for [100] uniaxial stress, allow to compute the band edges as follows:

\begin{equation}
\delta E_{g \Gamma}^{(2)} = \delta E_{g \Gamma} + a_{dir,100}^{(2)} \varepsilon_{100}^2
\end{equation}

\begin{equation}
\delta E_{g L}^{(2)} = \delta E_{g L} + a_{ind,100}^{(2)} \varepsilon_{100}^2
\end{equation}

Where $\delta E_{g \Gamma}$ (resp. $\delta E_{g L}$) are the direct (resp. indirect) band gaps calculated using the deformation potential theory (see supplementary information for detailed formulas).

In particular, since the heavy-hole band is fundamental for tensile uniaxial stress, we can express the relation between longitudinal strain and the lowest direct bandgap as:

\begin{equation}
E_{g\Gamma}(\varepsilon_{100}) \ (eV) = 0.80 - (7.31 \pm 0.15) \ \varepsilon_{100} - (37 \pm 7) \ \varepsilon_{100}^2
\end{equation}

Table \ref{tbl:defpot} summarizes the deformation potentials found in this work both experimentally via the electro-absorption technique, and theoretically, via tight-binding simulations, and compares them to previous works in the literature. The linear coefficients for the conduction bands as well as the tetragonal valence band shear deformation potential $b$ from the present work exhibit very good agreement with the already published theoretical and experimental data. Fitting the tight-binding results, we of course obtain similar coefficients as published in 2009 \cite{Niquet09}, but the introduction of a second order term slightly modifies the linear ones. We note however that the tight-binding $b$ deformation potential is slighly larger than experimental measurements, as it was adjusted on older data sets. In addition, the table shows the second order coefficients $a_{dir,100}^{(2)}$ and $a_{ind,100}^{(2)}$ we introduce to compute more precisely the evolution of the band gaps with the [100] uniaxial stress.

\renewcommand*{\thefootnote}{\alph{footnote}}

\begin{table}[ht]
\centering

\caption{\label{tbl:defpot}Deformation potentials for strained germanium from present work as compared to literature}

\begin{tabular}{p{3cm} p{3cm} p{2cm} p{3cm} p{3cm}}
\hline\hline
    & \multicolumn{2}{c}{Present work} & \multicolumn{2}{c}{Previous works} \\
		& Exp. & Theor. & Exp. & Theor. \\
\hline

$a_{c,dir} - a_v$                 & $-9.1 \pm 0.3$ & $-8.91$ &
$-8.97\pm0.16$\footnotemark[1]\newline
$-7.49$\footnotemark[2]\newline
$-10.9\pm0.8$\footnotemark[3]\newline
$-8.56\pm0.4$\footnotemark[4]\newline
$-9.77$\footnotemark[5]\newline
$-8.63$\footnotemark[6]\newline
$-10.9$\footnotemark[7]\newline
$-9.75$\footnotemark[18] &
$-9.48$\footnotemark[8]\newline
$-9.01$\footnotemark[9]\newline
$-9.13$\footnotemark[10]\newline
$-8.89$\footnotemark[11]\newline
$-8.77$\footnotemark[12]\newline
$-8.5$\footnotemark[13]\newline
$-8.8$\footnotemark[14]\newline
$-9.6$\footnotemark[15]\newline \\

$\Xi_d + \frac{1}{3} \Xi_u - a_v$ & --  & -3.20 &
$-2.0\pm0.5$\footnotemark[16]\newline
$-3.4\pm0.2$\footnotemark[17] &
$-2.78$\footnotemark[8]\newline
$-3.19$\footnotemark[9]\newline
$-2.71$\footnotemark[10]\newline
$-3.12$\footnotemark[12]\newline
$-2.6$\footnotemark[13]\newline
$-3.9$\footnotemark[14]\newline
$-3.5$\footnotemark[15]\newline \\

$b$                               & $-2.32 \pm 0.06$ & -2.72 &
$-1.88\pm0.12$\footnotemark[1]\newline
$-2.4\pm0.2$\footnotemark[2]\newline
$-2.86\pm0.15$\footnotemark[3]\newline
$-2.6\pm0.2$\footnotemark[4]\newline
$-1.88$\footnotemark[18] &
$-2.55$\footnotemark[8]\newline
$-2.74$\footnotemark[9]\newline
$-3.0$\footnotemark[13]\newline
$-3.1$\footnotemark[14]\newline
$-2.30$\footnotemark[15] \\

$a_{dir,100}^{(2)}$                 & $-37 \pm 7$ & -30 & -- & -- \\
$a_{ind,100}^{(2)}$                 & -- & -35 & -- & -- \\
\hline\hline
\end{tabular}

\begin{minipage}{2cm}
\footnotetext[1]{Ref. \cite{Liu04}}
\footnotetext[2]{Ref. \cite{Balslev67}}
\footnotetext[3]{Ref. \cite{Chandrasekhar77}}
\footnotetext[4]{Ref. \cite{Pollak68}}
\footnotetext[5]{Ref. \cite{Li94}}
\end{minipage}
\begin{minipage}{2cm}
\footnotetext[6]{Ref. \cite{Goni89}}
\footnotetext[7]{Ref. \cite{Welber77}}
\footnotetext[8]{Ref. \cite{VandeWalle89}}
\footnotetext[9]{Ref. \cite{Niquet09}}
\footnotetext[10]{Ref. \cite{Chang84}}
\end{minipage}
\begin{minipage}{2cm}

\footnotetext[11]{Ref. \cite{Rodriguez86}}
\footnotetext[12]{Ref. \cite{Alouani88}}
\footnotetext[13]{Ref. \cite{Klenner92}}
\footnotetext[14]{Ref. \cite{Blacha84}}
\footnotetext[15]{Ref. \cite{Schmid90}}
\end{minipage}
\begin{minipage}{2cm}
\footnotetext[16]{Ref. \cite{Balslev66}}
\footnotetext[17]{Ref. \cite{Ahmad89}}
\footnotetext[18]{Ref. \cite{Kurdi16}}
\end{minipage}
\end{table}

\section{Conclusion}

To conclude, we have measured the direct bandgap of germanium as a function of [100] uniaxial strain up to \unit{3.3}{\%} using electro-absorption spectroscopy. We demonstrate a non-linear relation between the direct band gap and the longitudinal strain. This shows that a first order theory using deformation potentials is not enough to reproduce the evolution of the band gaps of germanium at high strains. It would for instance lead to a \unit{50}{\milli\electronvolt} error for the light-hole band gap at \unit{3.3}{\%}, which corresponds to an error of about \unit{300}{\nano\meter} in wavelength.

More precise simulations such as tight-binding modelling are required to properly describe strain-induced band gap shifts for strain levels higher than \unit{2}{\%} where nonlinear effects have a preeminent role. 

Our experiments highlight the need for proper experimental verification of the dependence of the band structure on strain under the high strains attainable nowadays. The corrections will have a direct impact on the design of future optoelectronic and electronic devices using highly strained semiconductor layers.

\begin{acknowledgement}

This work was supported by the CEA projects `Phare DSM-DRT: Photonique' and `Operando' and has been performed with the help of `Plateforme technologique amont' in Grenoble.

\end{acknowledgement}

\section{Methods}

\subsection{Strained micro-bridges fabrication}

A backside polished Germanium-On-Insulator (GeOI) substrate (figure \ref{fgr:process}.a) obtained with the SmartCut\texttrademark{} process was used for the fabrication of the strained micro-bridges \cite{Reboud15}. These micro-bridges amplify the \unit{0.16}{\%} biaxial strain in  the germanium layer resulting from the difference between the thermal expansion coefficients of silicon and germanium \cite{Suess13}. The germanium layer was patterned with Reactive Ion Etching (RIE) using a recipe based on \ce{Cl2}, \ce{N2} and \ce{O2} and \unit{10}{\nano\meter} of aluminium as a hard mask (figure \ref{fgr:process}.b). The resulting micro-bridges are \unit{2}{\micro\meter} wide, \unit{800}{\nano\meter} thick and \unit{7}{\micro\meter} long. The membranes were then released by etching the silicon dioxide sacrificial layer in a Primaxx HF vapor reactor. The release of the structure induces the retractation of the stretching arms, which put the bridge under tensile strain \cite{Suess13}. The bridge geometry was chosen so as to obtain an homogeneous, pure tetragonal strain in the central region \cite{Tardif16}. The process used brought the whole released structure into contact with the silicon substrate, allowing further lithographic steps on the sample (figure \ref{fgr:process}.c).

\subsection{Spectroscopy}

Raman spectroscopy is performed to measure the strain in the micro-bridges. A \unit{785}{\nano\meter} laser with a \unit{9}{\micro\watt} output is used as a light source in a Renishaw inVia spectrometer. The resulting Raman spectrum is independent of the excitation power in this range. The Raman shift is then fitted by a Lorentzian curve and compared to the spectrum of bulk unstrained germanium. We calibrated the relationship between the Raman shift and the longitudinal uniaxial stretch along [100] using Laue X-ray micro-diffraction on selected micro-bridges (with strains up to \unit{4.9}{\%}) at the BM32 line of the European Synchrotron Radiation Facility (ESRF) \cite{Gassenq16-raman}. X-ray diffraction indeed provides a direct measurement of the lattice parameters of the micro-bridge, thus gives a model-free measurement of strain. The obtained relationship is then fitted with a second order polynomial and used to extract strains from Raman shifts for the samples of this study. No diffraction peak broadening was observed up to the most strained bridges, indicating the absence of plasticity (at the probe location). All shear components measured by X-ray diffraction are two orders of magnitude smaller than the diagonal components.

Electro-absorption experiments were performed with a Fourier Transform Infrared (FTIR) spectrometer using a quartz-tungsten-halogen bulb as a light source and a Cassegrain objective to focus the incident light onto a single micro-bridge. The transmitted light was collected on a Thorlabs extended \ce{InGaAs} photodiode and the generated signal was sent to a Stanford Research Systems SR830 lock-in amplifier. The electric field was applied by a Keysight 33500B function generator outputting a \unit{12}{\kilo\hertz} positive square wave and triggering the lock-in amplifier.

\clearpage

\bibliography{refs}

\end{document}